\documentclass[aps,pre,twocolumn,letter]{revtex4-1}
\usepackage{graphicx}
\usepackage{float}
\usepackage{amsmath}
\usepackage{color}
\usepackage{xr}

\begin{document}
\title{Geometric order parameters derived from the Voronoi tessellation show signatures of the jamming transition}

\date{\today}
\author{Peter K. Morse and Eric I. Corwin}
\affiliation{Department of Physics and Materials Science Institute, University of Oregon, Eugene, Oregon 97403, USA.}

\begin{abstract}
A jammed packing of frictionless spheres at zero temperature is perfectly specified by the network of contact forces from which mechanical properties can be derived.  However, we can alternatively consider a packing as a geometric structure, characterized by a Voronoi tessellation which encodes the local environment around each particle.  We find that this local environment characterizes systems both above and below jamming and changes markedly at the transition.  A variety of order parameters derived from this tessellation carry signatures of the jamming transition, complete with scaling exponents.  Furthermore, we define a real space geometric correlation function which also displays a signature of jamming.  Taken together, these results demonstrate the validity and usefulness of a purely geometric approach to jamming.

\end{abstract}

\maketitle

\section{Introduction}

Over the past two decades the jamming of athermal frictionless spheres has been seen as the limiting case of several different kinds of systems\cite{liu_nonlinear_1998, ohern_random_2002, mari_jamming_2009, parisi_mean-field_2010, bi_jamming_2011, charbonneau_jamming_2015}. Athermal soft sphere systems can be brought to the limit of zero internal energy and isostaticity, achieving a critically jammed system which is typically characterized by mechanical properties \cite{ohern_random_2002, ohern_jamming_2003, wyart_rigidity_2005, hecke_jamming_2010, tighe_model_2010, bi_jamming_2011}.  However, when such systems are below the jamming density there is no longer a mechanical network of force-bearing contacts and so mechanical order parameters are all identically zero.  Conversely, hard sphere thermal liquids are studied below the glass or jamming transition and are characterized by dynamic quantities such as mobility and pressure \cite{whitelam_dynamic_2004, parisi_mean-field_2010, charbonneau_jamming_2015}.  As density is increased they reach the limit of diverging reduced pressure and become a critically jammed system.  Above this density, hard sphere systems can not exist.  While both athermal soft sphere systems and thermal hard sphere glass systems have been successful models for predicting and measuring scaling exponents of various parameters near the jamming phase transition \cite{ohern_random_2002, tighe_model_2010, parisi_mean-field_2010, charbonneau_jamming_2015}, neither of these model systems speak to the behavior of unjammed athermal systems.  This leaves a gap in the understanding of the athermal jamming transition. In this paper we introduce new geometric order parameters which characterize the athermal jamming transition both above and below jamming, placing both sides of the transition on equal footing and providing a meaningful way to interrogate soft sphere systems below the jamming transition.

The structure of jammed systems has long been studied in terms of geometry \cite{meijering_interface_1953, bernal_random_1967, gotoh_statistical_1974, dodds_simplest_1975, slotterback_correlation_2008, clusel_granocentric_2009} however a systematic study of geometric changes as a function of distance to the transition has not yet been performed. The Voronoi tessellation \cite{voronoi_nouvelles_1908}, which is well defined at all packing fractions, provides a natural lens through which to study both unjammed and overjammed systems.  In previous work we have demonstrated that the number of facets (corresponding to the number of nearest neighbors) provides a good order parameter for the jamming transition \cite{morse_geometric_2014}.  This order parameter raised a new problem, however, because it exhibited an upper critical dimension (above which, all order parameters share the same scaling laws) of $d=3$. This stood in contrast to the well known fact that mechanical order parameters exhibit an upper critical dimension of $d=2$ \cite{ohern_random_2002,goodrich_finite-size_2012}. This, coupled with the recent success of replica theory in predicting high finite dimensional scaling \cite{charbonneau_jamming_2015} has motivated us to explore a range of geometric order parameters in dimensions ranging from $d=2$ to $d=5$. 

In this paper we show that most geometric properties of the Voronoi tessellation are controlled by the jamming point $\phi_J$, suggesting that jamming can be described in purely geometric terms. Further, we present a new geometrically defined correlation function which changes qualitatively at the jamming transition.  Surprisingly, none of these measures show any indication of the previously discovered pre-jamming transition, associated with the maximum inscribed sphere of the Voronoi cell, which we have found to happen at a density $\phi^* < \phi_J$ \cite{morse_geometric_2014}.

\begin{figure}[p]
\includegraphics [width=.9\linewidth]{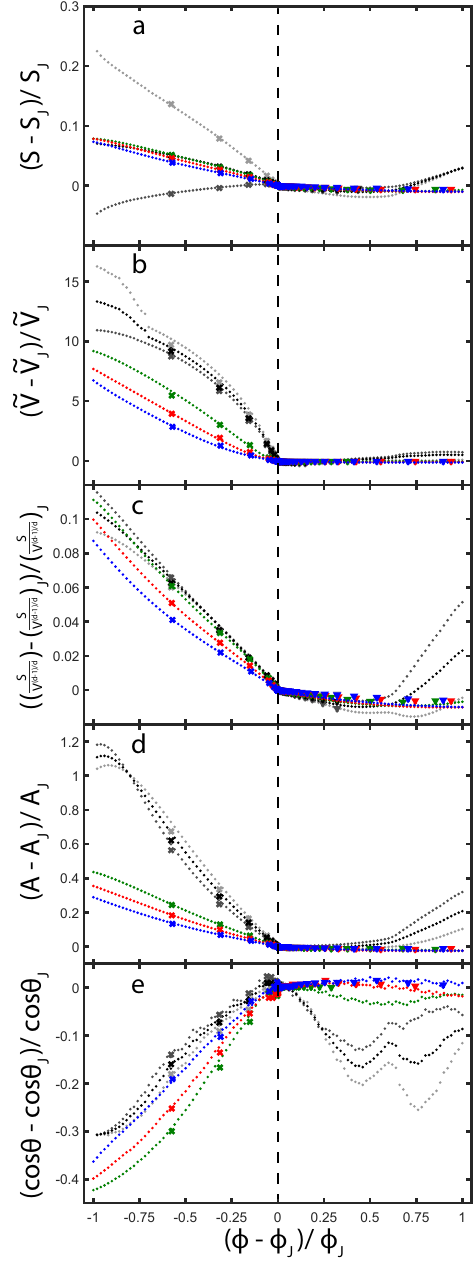}
\caption{Plots of scaled order parameters vs. the scaled packing fraction. Closed circles represent IQ data, x's represent GM data (from below), and triangles represent ES data (from above). The parameters shown are (a) mean surface area, $S$, (b) standard deviation of volume divided by the mean of the volume, $\tilde{V}$, (c) mean surface to volume ratio $S/V^{(d-1)/d}$ (d) mean aspect ratio, $A$, and (e) mean aspect ratio angle $cos\theta$. We plot data for $d=2$ (smaller particles light gray, larger particles dark gray, combined black), $d=3$ (green), $d=4$ (red), and $d=5$ (blue).}
\label{fig:newIQ}
\end{figure}


\begin{figure*}[p]
\includegraphics [width=.90\textwidth]{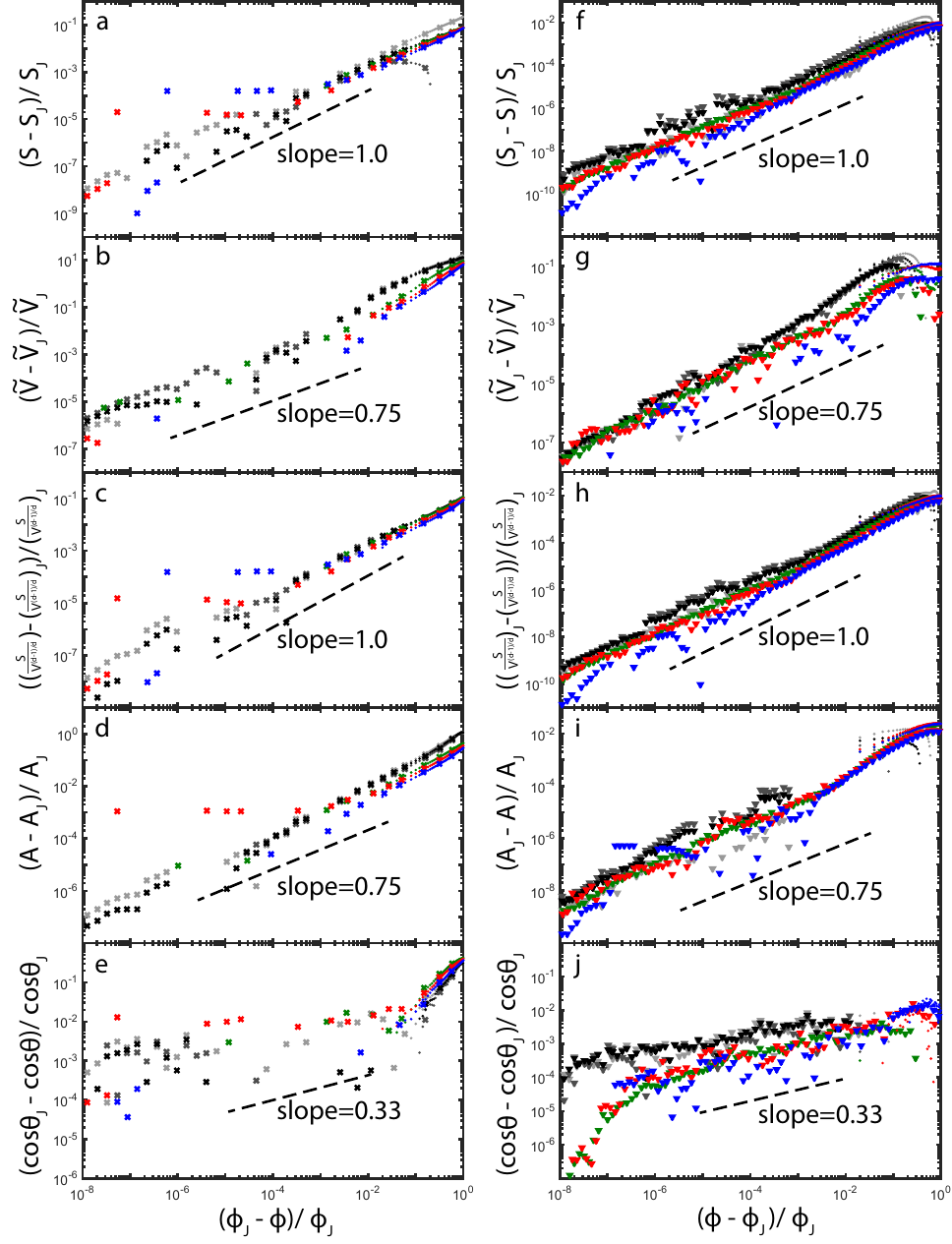}
\caption{Log-log plots of each scaled order parameter vs. the scaled packing fraction approaching jamming from below (left) and above (right). Closed circles represent IQ data, x's represent GM data (from below), and triangles represent ES data (from above). The parameters shown are (a,f) mean surface area, $S$, (b,g) standard deviation of volume divided by the mean of the volume, $\tilde{V}$, (c,h) mean surface to volume ratio $S/V^{(d-1)/d}$ (d,i) mean aspect ratio, $A$, and (e,j) mean aspect ratio angle $cos\theta$. We plot data for $d=2$ (smaller particles light gray, larger particles dark gray, combined black), $d=3$ (green), $d=4$ (red), and $d=5$ (blue).}
\label{fig:newPLBoth}
\end{figure*}

\section{Generating a packing}

We simulate packings of frictionless athermal particles with a harmonic contact potential in periodic boundary conditions as described in references \cite{charbonneau_universal_2012, morse_geometric_2014}. In $d=3-5$, we use monodisperse spheres, and in two dimensional systems, we use a 50:50 mixture of bidisperse disks with a ratio of radii that is 1:1.4, known to show mechanical jamming. We present data obtained with three packing protocols: Infinite Quench (IQ)\cite{ohern_random_2002}, Geometric Mean (GM)\cite{charbonneau_universal_2012, morse_geometric_2014}, and Energy Sweep (ES)\cite{charbonneau_jamming_2015}.  

Our three protocols differ only in how jamming is approached. All begin with a set of particles in random positions at a specified density.  The energy of this system is then minimized to find the so-called inherent structure, found at the local energy minimum.  Each of these protocols works as an iterative process by finding the inherent structure at a given density and then using this packing as the seed to find a minimized packing at a new density.

The \textbf{IQ protocol} begins with a random packing at zero density.  Every particle is inflated to achieve a new packing at a specified higher density and this packing’s energy is then minimized.  The results of this minimization are then used to create a denser packing, and so on.  This proceeds in linearly spaced steps of packing fraction until the desired range of packing densities is covered.  The range is chosen to cover densities from $\phi=0$ to $\phi=2\phi_J$.  The limits of this range are somewhat arbitrary but are chosen to be symmetric about $\phi_J$. While the most relevant region is near the transition point, we include data at both the high and low extremes for completeness. Data for $d=3-5$ uses 65536 ($2^{16}$) particles, while $d=2$ uses 16384 ($2^{14}$) particles.

The \textbf{GM protocol} is designed to zero in on the transition point, either approaching from above or below, without ever overshooting. In this manuscript, we only report on GM systems approaching from below because the ES protocol (described below) converges much faster when approaching from above. The GM protocol requires an initial bounding of the jamming point by choosing two densities, one above and one below. A packing is initially created at the lower bound and its energy minimized.  A new packing is created between the upper and lower bounds using the original packing as its seed.  If this packing is below jamming (taken to mean an energy per particle of $<10^{-20}$), it becomes the new lower bound and serves as the next seed.  If, however, this packing is found to be above jamming it is discarded and its density is used as the upper bound in picking a new intermediate density.  This proceeds until we approach the jamming point to within our energy per particle tolerance of $10^{-20}$.  In this way we are able to create a packing right at the edge of jamming that is the result of only inflationary steps, without ever crossing into the jammed regime. Because the convergence is slow, we are only able to report on 8192 ($2^{13}$) particles.

The \textbf{ES protocol} is limited in that it can only serve to approach jamming from above, but as previously mentioned, it converges faster than GM. The ES protocol exploits the scaling of system energy with excess packing fraction $E \propto \left( \phi-\phi_J \right)^2$ to gently approach jamming from above, creating $n_\textrm{steps}$ logarithmically spaced packings per decade.  Given an initial system density $\phi_i$, system energy $E_i$, and a guess for the jamming density $\tilde \phi_i$ we calculate the packing fraction for the next system as
\begin{align}
  \phi_{i+1} & = \tilde\phi_i + \left( \phi_i - \tilde\phi_i \right) 10^{-1/n_\textrm{steps}}.
\end{align}
Once this new system's energy is minimized we compute a better estimate for the true jamming density as
\begin{align}
  \tilde\phi_{i+1} & = \frac{\phi_{i+1} - \phi_i \sqrt{ E_i/E_{i-1} } }{1 - \sqrt{ E_i/E_{i-1} } }.
\end{align}
This process continues until we achieve an energy per particle of $10^{-20}$.

We choose the starting point of the approach to be approximately $2\phi_J$.  It has been previously shown that the jamming density when approaching from above is dependent on the initial packing density for systems that start close to $phi_J$ \cite{charbonneau_universal_2012}.  We choose to start at such a high value of $\phi$ to ensure that our results are independent of the starting density.

All ES data sets use 16384 ($2^{14}$) particles. Data for $d=2$, $d=3$, and $d=4$ are averaged over 10, 63, and 79 systems respectively while data for $d=5$ is taken from a single run.

\section{Geometry of the Voronoi tessellation}

\begin{figure}[h]
\includegraphics [width=.6\linewidth]{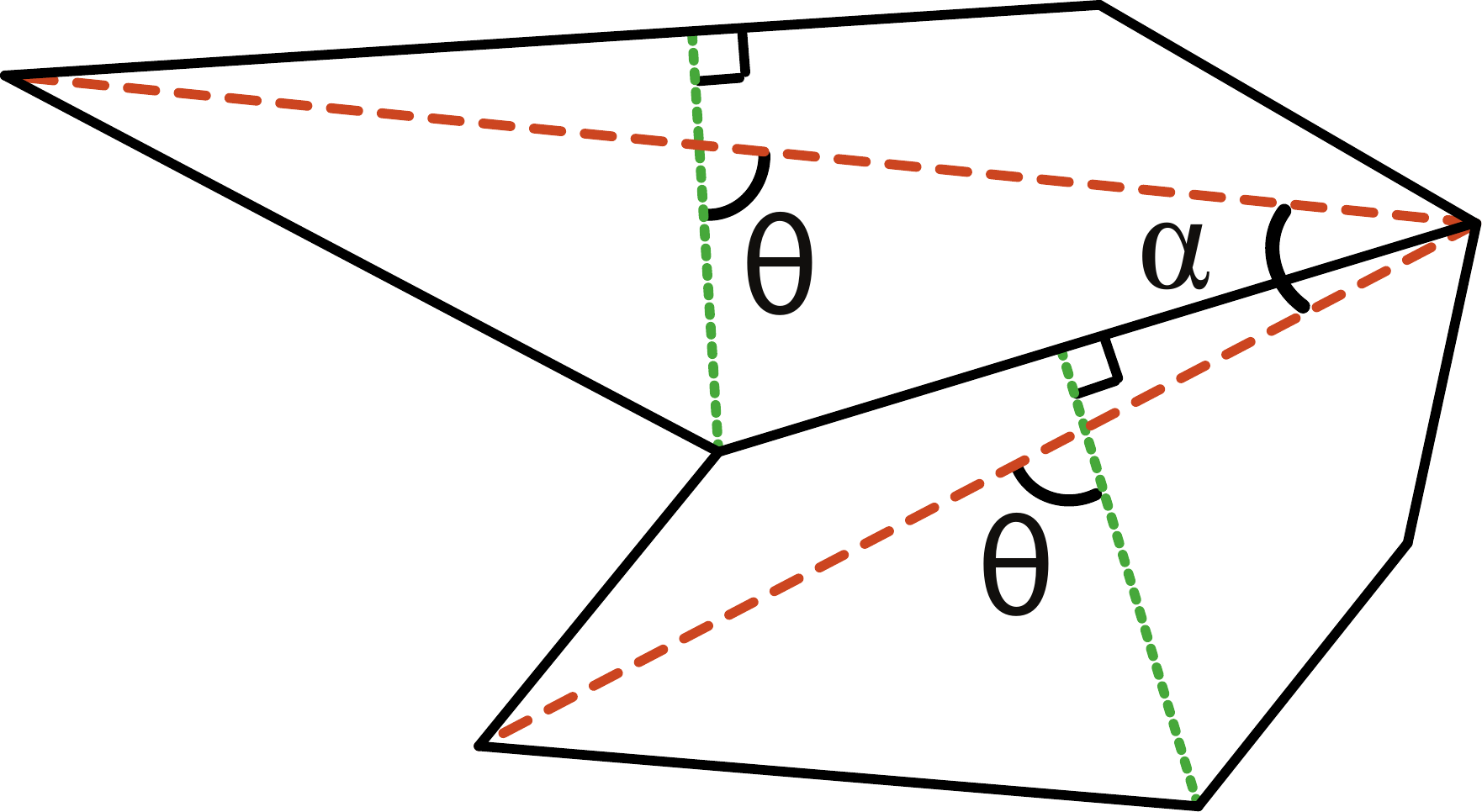}
\caption{Illustration of the aspect ratio axes in two Voronoi cells. For each cell, the short axis is shown in green (short dashes) and the long axis is shown in orange (long dashes). The angle $\theta$ between the two axes is defined to be the acute angle between the short and long axis. The angle $\alpha$ between two long axes of different cells is also shown.}
\label{fig:aspectRatioDiagram}
\end{figure}

Given a packing created via any of our protocols and in any dimension we calculate the Voronoi tessellation using the algorithms described in \cite{morse_geometric_2014} and extract the associated vertices using the Delaunay triangulation \cite{delaunay_sur_1934}.  For the monodisperse packings we create in $d=3-5$, this Voronoi tessellation is the standard Voronoi tessellation wherein the size of a cell is independent of the size of the particle. However, due to the bidispersity used in $d=2$ we use the radical Voronoi (or Laguerre) tessellation \cite{voronoi_nouvelles_1908}, which makes the boundaries between cells the bisecting plane between the particle edges. This preserves the convexity of each cell and thus provides a natural extension of the classical Voronoi cell. From each Voronoi cell, we extract all of our measurements. The number of facets of the Voronoi tessellation gives us 1) the number of nearest neighbors $N$; The vertices of the Voronoi cell allow us to calculate 2) the surface area $S$ and 3) the volume $V$; The ratio between the largest and smallest possible distances between parallel planes kissing the cell defines 4) the aspect ratio $A$; Finally, the dot product between the headless vectors defining the aspect ratio provides 5) the cosine of the cell's internal angle $\theta$.

\subsection{Volume and Surface Area}

\begin{table}
\begin{tabular}{l|c|c|c|c|c|c}
Parameter $\chi$ & $N$ & $S$ & $\tilde V$ & $S/V^{(d-1)/d}$ &  $A$ &  $\cos\theta$\\
\hline
Power $\gamma$ & 0.7 & 1.0 & 0.75 & 1.0 & 0.75 & 0.33\\
\hline
$\chi_J$, $d=2$, large & $6$ & $3.005$ & $0.0315$ & $3.738$ & $1.221$ & $0.567$\\
\hline
$\chi_J$, $d=2$, small & $6$ & $2.275$ & $0.0478$ & $3.827$ & $1.306$ & $0.559$\\
\hline
$\chi_J$, $d=2$, all & $6$ & $2.640$ & $0.0573$ & $3.782$ & $1.264$ & $0.563$\\
\hline
$\chi_J$, $d=3$ & $14.29$ & $5.385$ & $0.0386$ & $5.386$ & $1.322$ & $0.429$\\
\hline
$\chi_J$, $d=4$ & $32.74$ & $6.874$ & $0.0366$ & $6.875$ & $1.379$ & $0.385$\\
\hline
$\chi_J$, $d=5$ & $74.62$ & $8.261$ & $0.0340$ & $8.262$ & $1.412$ & $0.350$\\
\end{tabular}
\caption{Scaling laws and critical values for all parameters $\chi$, such that $\frac{\chi-\chi_J}{\chi_J} \propto (\frac{\phi - \phi_J}{\phi_J})^\gamma$. All critical values are unitless except for $S_J$ which is reported as the unitless $S_JN_\textrm{particles}^{(d-1)/d}$. For $d=2$, we report separately on $\chi_J$ values for the larger particles, the smaller particles, and the system as a whole.}
\label{tab:Scaling}
\end{table}

Calculating volumes and surface areas is notationally complicated but conceptually simple to achieve by breaking the cell into simplices.  To find volumes and surface areas we exploit the fact that the $d$-dimensional volume of a $d$-simplex can be calculated from the generalized triple product of its vertices.  The Delaunay triangulation of the surface of a Voronoi cell breaks down the surface of each facet $k$ into a number of $(d-1)$ dimensional simplices labeled by the index $m$. There are $d$-vertices association with each simplex, which we denote as  $\vec{v}_{m,i}$ where $i$ ranges from $1$ to $d$, and we denote the outward facing normal vector to a facet $k$ as $\hat{n}_k$. From this, the surface area of each facet is calculated as the sum of the surface of all of its constituent simplices as

\begin{align}
S_k & = \sum_m \frac{|\hat{n}_k \cdot \big[ (\vec{v}_{m,1} - \vec{v}_{m,d}) \wedge \cdots \wedge (\vec{v}_{m,d-1} - \vec{v}_{m,d}) \big]| } {(d-1)!},
\end{align}

where $\wedge$ denotes the $d$-dimensional wedge product. The total surface area of a given Voronoi cell is then the sum of all facets

\begin{align}
S & = \sum_k S_k.
\end{align}

By choosing an interior point of the cell $\vec{r}$, we can subdivide the volume of the cell into a number of $d$-simplices whose volumes sum to the volume of the cell as


\begin{align}
V =  \sum_m \frac{|(\vec{v}_{m,d} - \vec{r}) \cdot \big[ (\vec{v}_{m,1} - \vec{v}_{m,d}) \wedge \cdots \wedge (\vec{v}_{m,d-1} - \vec{v}_{m,d}) \big]| }{d !}.
\end{align}

For a given packing, the mean cell volume is just the simulation volume divided by the number of particles. The distribution of cell volumes, however, does change, and so we report on $\tilde V$, the ratio of the standard deviation of the volume distribution to the mean. We also report on the mean of the unitless surface to volume ratio $S/V^{(d-1)/d}$.

\begin{figure}[p]
\includegraphics [width=.90\linewidth]{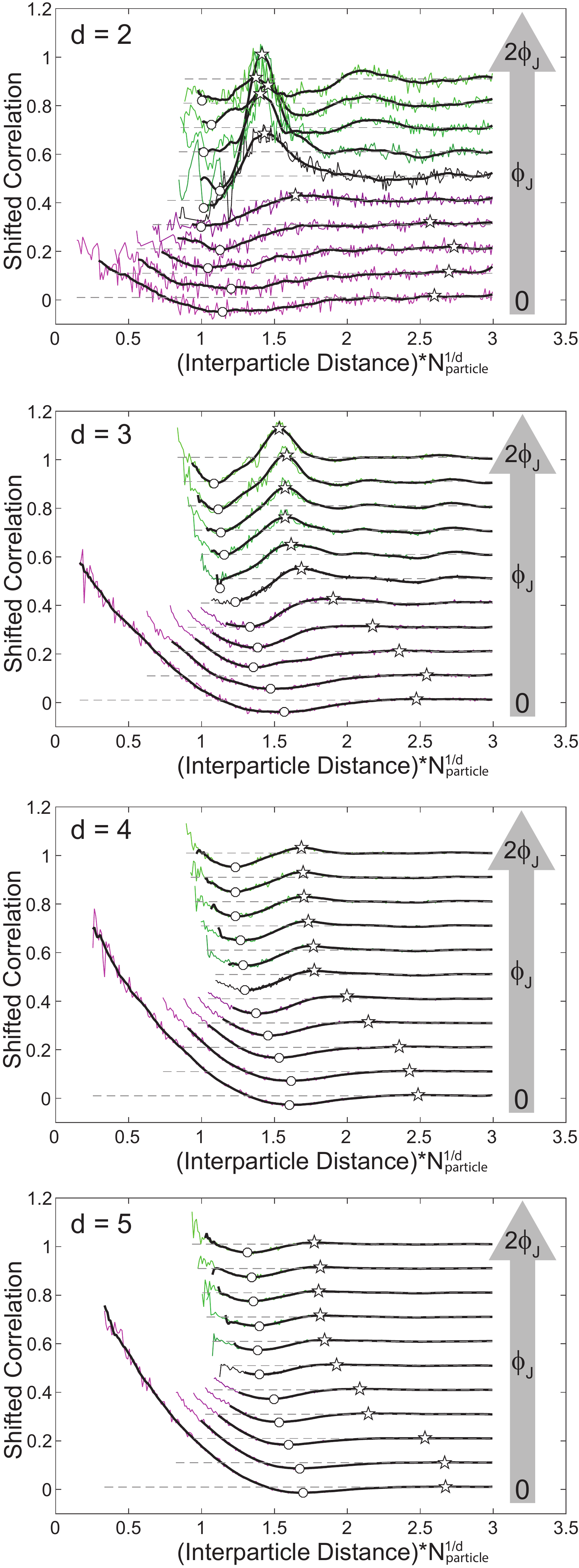}
\caption{The normalized long axis correlation between Voronoi cells plotted as a function of distance between particles in $d = 2-5$. Correlations are shifted vertically to show the effect of changing $\phi$ with a color scheme that fades from purple ($\phi = 0$) to black ($\phi = \phi_J$) to green ($\phi = 2\phi_J$). Gray dashed lines show the line corresponding to completely uncorrelated axes, open circles denote minima and open stars represent secondary maxima. A black line has been drawn over each curve representing the Savitsky-Golay filter. Data obtained using the IQ protocol.}
\label{fig:correlationPlotAll}
\end{figure}

\subsection{Aspect Ratio and Internal Angle $\theta$}

The ratio of surface area to volume $S/V^\frac{d-1}{d}$ defines a simple notion of an aspect ratio, but one that is insensitive to the anisotropy of the cell.  We define another aspect ratio, explicitly sensitive to anisotropy by looking at the ratio between the longest one dimensional span in a cell to the shortest one dimensional span of a cell (Figure \ref{fig:aspectRatioDiagram}). To calculate this aspect ratio we define the long axis $\vec \ell$ as the maximum distance between any pair of vertices and the short axis $\vec s$ as the minimum of the set of maximum distances between each vertex and each facet.  Given a set of vertices $\vec{v_i}$ and introducing a point $\vec{p}_k$ on each facet $k$, these definitions can be formalized as


\begin{align}
\vec{\ell} & = \{ \vec{\ell}\, \mid \|\vec{\ell}\| = \textrm{Max}_{ij} \|\vec{v}_i - \vec{v}_j\| \},
\end{align}
and 
\begin{align}
\vec{s} &= \{ \vec{s} \, \mid \|\vec{s}\| = \textrm{Min}_k(\textrm{Max}_i|\hat{n}_k \cdot (\vec{v}_i - \vec{p}_k)|)\}.
\end{align}

The aspect ratio is then simply defined as

\begin{align}
A & = \frac{\| \vec\ell \|}{\| \vec s \|}.
\end{align}

We can further measure the skewness of a cell by defining the angle between the long axis and the short axis as
\begin{align}
\cos \theta = \frac{|\vec{\ell}\cdot\vec{s}|}{\|\vec{\ell}\|\|\vec{s}\|},
\end{align}
where the absolute value is taken because these are headless vectors.
\subsection{Correlation Function}

We can examine the interaction of each cell with its neighbors by defining a correlation function based on the angle between the axes of pairs of cells.  When cells are packed together to fill space neighboring cells must share facets, potentially causing the axes to align.  To characterize this we measure the cosine of the angle between two long axes $\vec{\ell}_i$ and $\vec{\ell}_j$ associated with particles $i$ and $j$ respectively (illustrated in Figure \ref{fig:aspectRatioDiagram}). Because the axes are headless vectors we must use the formalism of directors, giving rise to the definition for the cosine as
\begin{align}
\cos \alpha_{ij} = \frac{| \vec{\ell}_i \cdot \vec{\ell}_j |}{\|\vec{\ell}_i\| \|\vec{\ell}_j \|}.
\end{align}

To compare systems in different dimensions, we must first calculate the expectation values of completely uncorrelated directors. The expectation value of the cosine of the angle in dimension $d$ is given by

\begin{align}
\langle \cos \alpha \rangle_d & = \frac{\int_{0}^{\pi/2} \cos \alpha \sin ^{d-2} \alpha d\alpha}{\int_{0}^{\pi/2} \sin^{d-2} d \alpha}  = \frac{\Gamma\left(\frac{d}{2}\right)}{\sqrt{\pi}\, \Gamma\left(\frac{d+1}{2}\right)}.
\end{align}
The standard deviation of the angle between uncorrelated directors in dimension $d$ is defined as $\sigma_d = \sqrt{\langle \cos \alpha \rangle_d^2 - \langle \cos^2 \alpha \rangle_d}$.  Therefore we also calculate the expectation of the square of the cosine of the angle of uncorrelated directors as
\begin{align}
\langle \cos^2 \alpha \rangle_d & = \frac{\int_{0}^{\pi/2} \cos^2 \alpha \sin ^{d-2} \alpha d\alpha}{\int_{0}^{\pi/2} \sin^{d-2} d \alpha} & = \frac{\Gamma\left(\frac{d}{2}\right)}{2\,\Gamma\left(d+\frac{1}{2}\right)}.
\end{align}
Thus we find the standard deviation of uncorrelated directors in dimension $d$ to be
\begin{align}
\sigma_d & =  \sqrt{\frac{\Gamma\left(\frac{d}{2}\right)^2}{\pi\, \Gamma\left(\frac{d+1}{2}\right)^2} - \frac{\Gamma\left(\frac{d}{2}\right)}{2\,\Gamma\left(d+\frac{1}{2}\right)}}.
\end{align}

We define our correlation function as the normalized value of the cosine of the angle between the long axes of every pair of particles as a function of the distance between cells as
\begin{align}
C_{\ell}(r) & = \sum_{ij} \delta(\|\vec{r}_i - \vec{r}_j\| - r)\,\frac{\cos \alpha_{ij} - \langle \cos \alpha \rangle_d }{\sigma(d)}.
\end{align}

We note that this correlation function, relating the shape and asymmetry of Voronoi cells, is logically distinct from the pair correlation function, or indeed from any correlation function based solely on particle positions.

\section{Analysis}

\subsection{Order Parameters}

Figure \ref{fig:newIQ} presents the geometric order parameters described above calculated for systems created with all protocols as a function of distance to $\phi_J$ for $d=2-5$.  Data across multiple dimensions is presented on the same scale by subtracting off the value at the jamming transition and then dividing by that same value.  The packing fraction is similarly scaled as $(\phi - \phi_J)/\phi_J$.  Below jamming all of these parameters change rapidly with increasing packing fraction.  Jamming is marked by a sharp kink and above jamming they evolve with a much gentler slope. For all measures except the surface area to volume ratio the $d=2$ data does not seem to collapse onto the same family of curves as the higher dimensions.  Because the $d=2$ packings are bidisperse we show separate curves for each particle size (shown in dark and light gray) and a single curve representing the combined data (shown in black). The difference is especially apparent in the surface area: the Voronoi surface area increases for larger particles and decreases for smaller particles as jamming is approached from below. This makes intuitive sense; at extremely low packing fractions the Voronoi cells for the two sets of particles should be almost identical and near jamming the larger particles will end up having a larger surface area and a larger volume than their smaller counterparts. In the combined data, the curve collapses to follow the trend observed in $d=3-5$.

In order to explore the behavior very close to jamming we use the GM protocol to approach from below and ES to approach from above. In this way we obtain packings that converge logarithmically on $\phi_J$. Plotted on a log-log scale (Figure \ref{fig:newPLBoth}) we find that each parameter scales with its own power-law on both sides of the transition with power law values and critical values listed in Table \ref{tab:Scaling}. We have previously demonstrated that the mean number of neighbors $\langle N \rangle$ scaling is consistent with a power of $\sim$0.7 \cite{morse_geometric_2014}.

Below jamming, there must be a limit to the scaling regime.  The mean surface area $\langle S \rangle$, volume $\langle V \rangle$, and number of facets $\langle N \rangle $ of Voronoi cells at $\phi=0$ and their respective dimensional dependence can be semi-analytically determined \cite{meijering_interface_1953, moller_random_1992, pineda_domain-size_2007, farjas_cell_2008}. The same should be true for aspect ratio $\langle A \rangle$ and internal angle $\langle \cos\theta \rangle$ but to our knowledge those studies have not yet been done.  This is responsible for the changes in curvature seen at low $\phi$ in Figure \ref{fig:newIQ}.

While most of the power laws work well over at least five decades, there are a few exceptions. The data from below is very sparse, and so we cannot claim that the power laws fit exactly and can only suggest that the plots look like power-laws within the plotting area. Precise claims about the scaling exponents of these power laws would require a method which approached jamming more predictably from below and which converged much faster so that averaging could be used, as it is done above jamming.

It is also important to note that the $d=2$ data deviates significantly in the standard deviation of the volume (Figure \ref{fig:newPLBoth}g) and the internal angle (Figure \ref{fig:newPLBoth}j). When coupled with the fact that the mean number of neighbors does not show a signature of jamming in $d=2$ \cite{morse_geometric_2014}, this strongly suggests that $d=2$ is below the upper critical dimension of the jamming transition when viewed from a geometric perspective.

\begin{figure}[h]
\includegraphics [width=.90\linewidth]{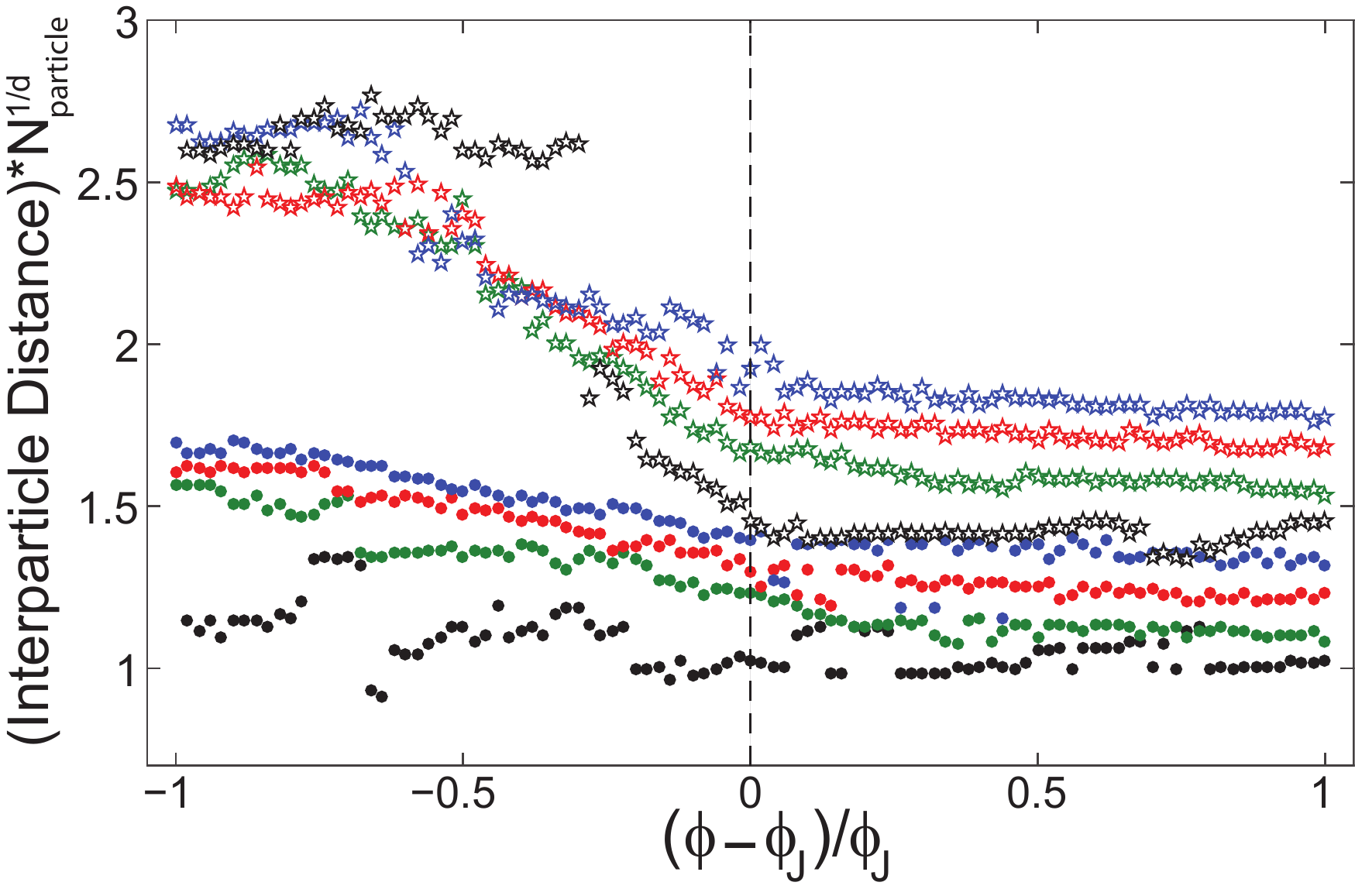}
\caption{The position of the minimum (closed circles) and secondary maximum (open star) of the correlation shown in Figure \label{fig:aspectRatio} as a function of distance from the jamming transition. Colors shown represent dimensions 2 (black), 3 (green), 4 (red), and 5 (blue).}
\label{fig:correlationOverPhi}
\end{figure}

\subsection{Correlation Function}

From the measurements of the aspect ratio we can see that at jamming the Voronoi cells are much more isotropic than they are far from jamming.  At jamming, the aspect ratio is close to 1 and the direction of the long and short axes are uncorrelated as measured by $\cos \theta$.  In contrast, at very low density the cells are elongated and have a large aspect ratio and axes that are nearly perpendicular.  Figure \ref{fig:correlationPlotAll} shows the measured correlation function between the long axes as a function of interparticle distance for packing fractions ranging from $\phi=0$ to $\phi=2 \phi_J$ in dimensions $d=2-5$. Far below jamming, neighboring particles are highly correlated.  This correlation decreases with increasing distance, showing an anti-correlated dip at intermediate distances and then finally decaying to completely decorrelated at large distances.  At jamming, the correlation function changes qualitatively, marked by the appearance of a positive correlation peak at intermediate distances in addition to the short distance dip.  Both the dip and the peak become more prominent and sharpen at higher packing fractions.  These extrema are found using a cubic Savitzky-Golay filter with a span of 51 data points \cite{savitzky_smoothing_1964} and the positions of the dip and peak are indicated by circles and stars respectively in Figure \ref{fig:correlationPlotAll} and plotted as a function of packing fraction in Figure \ref{fig:correlationOverPhi}. The position of the maximum shows a clear signature of the transition in $d=2-5$.  However, the position of the minimum for $d=3-5$ does not show a clear signature of this transition. We find that the correlation functions plotted in Figure \ref{fig:correlationOverPhi} only depend on interparticle distance, with no angular dependence when oriented to the long axes of each given particle.  This correlation function is unusual in that the jamming transition is marked by the disappearance of the nearest neighbor correlation, seen in the value of the correlation function at the shortest possible interparticle distance.

\section{Conclusion}

We have observed a clear signature of the jamming transition in each of the studied measures of the Voronoi cell as well as in our newly defined axis-correlation function.  These results bolsters the claim that while jamming is a mechanical transition, it can be viewed separately as a purely geometric phenomenon. These results justify the use of the Voronoi cell as a tool to understand the jamming transition. Furthermore, we provide evidence that while the mechanical transition has an upper critical dimension of $d=2$, the geometric transition has an upper critical dimension of $d=3$ when considering some geometric order parameters.

Ultimately, each of the measures are sensitive to the fluctuations in the size and shape of individual Voronoi cells. Each measure reflects a different change in the cell.  The fact that we see power-law scaling in all of these measurements suggests that nearly every aspect of the cell changes and is controlled by the transition from unjammed to jammed.   Our results demonstrate that the mechanical jamming transition coincides perfectly with a transition in the geometry of the packing at $\phi_J$.


\begin{acknowledgments}
We thank John Royer for helpful discussions.  We thank the NSF for support under Career Award DMR-1255370. The ACISS supercomputer is supported under a Major Research Instrumentation grant, Office of Cyber Infrastructure, OCI-0960354.

\end{acknowledgments}

\bibliography{GeoScaling}

\end{document}